\input harvmac.tex
\input epsf


\lref\itdr{Itzykson, C. and Drouffe, J.-M.:
Statistical Field Theory. Vol.1:
Cambridge University Press (1989)}

\lref\coloda{Zamolodchikov, Al.B.:
Two-point correlation function
in scaling Lee-Yang model.
Nucl. Phys. {\bf B348}, pp. 619-641 (1991)\semi
Acerbi C., Mussardo G., and Valleriani, A.:
Form-factors and 
correlation functions of the stress - energy tensor in massive
deformation 
of the minimal models $E_{(N)}^{(1)}\times E_{(N)}^{(1)}/E_{(N)}^{(2)}$.
Int. J. Mod. Phys. {\bf A11}, pp. 5327-5364 (1996)\semi
Balog, J. and  Niedermaier, M.:
Off-shell dynamics of the 
O(3) NLS model beyond monte carlo and perturbation theory.
Nucl. Phys. B{\bf 500}, pp. 421-461 (1997) }

\lref\oldref{ Vergeles, S. and Gryanik., V.: Two-dimensional
Quantum Field Theories having exact solutions.
Yadern. Fiz. {\bf 23}, pp. 1324-1334 (1976)\ (in
Russian)\semi
Karowski,  M., Weisz, P.:
Exact form factors in (1+1)-dimensional
field theoretic models with solution behavior.
Nucl. Phys. {\bf B139}, pp. 455-476 (1978)\semi
Zamolodchickov, A.B.: Quantum sine-Gordon model. 
The soliton form-factors.  ITEP-45-1977,
Preprint 
ITEP,  12pp. (1977) }

\lref\drew{
Babujian, H., Fring, A.,  Karowski, M. and  Zapletal, A.:
Exact form-factors 
in integrable quantum field theories: the sine-Gordon model.
Nucl. Phys. {\bf B538}, pp. 535-586 (1999)\semi
Delfino, G.: Off critical correlations in the Ashkin-Teller model.
Phys. Lett. {\bf B450}, pp. 196-201 (1999)   }

\lref\Korep{Korepin, V.E. and Essler, F.H.L.:
$SU(2)\otimes SU(2)$ invariant scattering matrix of the Hubbard model.
Nucl. Phys. {\bf B426}, pp. 505-533 (1994)} 

\lref\ZZ{Zamolodchikov, A.B. and Zamolodchikov, Al.B.:
Structure Constants and Conformal Bootstrap in
Liouville Field Theory.
Nucl. Phys. {\bf B477}, 577-605   (1996) }

\lref\Baxter{Baxter, R.J.: Exactly Solved Models
in Statistical Mechanics. London: Academic Press (1982)}

\lref\Woyn{ 
Woynarovich F.: Excitations with 
complex wave numbers in a Hubbard chain. 1. States with one pair of
complex wave numbers. J. Phys. {\bf C}16, pp. 5293 (1983)\semi
Excitations with 
complex wave numbers in a Hubbard chain. 2. States with 
several pairs
of complex wave numbers. 
J. Phys. {\bf C}16, pp. 6593 (1983)}

\lref\lutem{Emery, V.J., Luther, A.  and Peschel, I.:
Solution of the one-dimensional electron gas on a lattice.
Phys. Rev. {\bf B13}, pp.1272-1276 (1976)}
 
\lref\fil{Filev, V.M.: 
Spectrum of two-dimensional relativistic model.
Teor. i Mat. Fiz. {\bf 33}, pp. 119-124 (1977)}

\lref\Mel{Melzer, E.:
On the scaling limit of the 1-d Hubbard model at half filling.
Nucl. Phys. {\bf B443}  [FS], pp. 553-564 (1995)}

\lref\Woynarovich{Woynarovich, F. and  Forgacs, P.:
Scaling limit of the one-dimensional attractive Hubbard model: 
The half-filled band case.
Nucl. Phys. {\bf B498}, pp.65-603 (1997)}

\lref\Jons{Johnson, J.D., Krinsky, S. and McCoy, B.M.:
Vertical-arrow correlation length in
the eight-vertex model and the low-lying excitation
of the X-Y-Z Hamiltonian.
Phys. Rev. {\bf A8}, 2526-2547 (1973)}

\lref\luth{Luther, A.: Eigenvalue
spectrum of interacting massive fermions in one dimension.
Phys. Rev. {\bf B14}, 5, 2153-2159 (1976)}

\lref\BeLeC{Bernard, D. and LeClair, A.:
Residual quantum symmetries of
the restricted Sine-Gordon theories.
Nucl. Phys. {\bf B340},
pp. 721-751 (1990)}

\lref\miw{Foda, O.,  Iohara, K., Jimbo, M., Kedem R., Miwa, T. and
Yan, H.: An elliptic quantum algebra for SL(2).
Lett. Math. Phys. {\bf 32}, pp. 259-268 (1994)}

\lref\LYa{Lukyanov, S. and Pugai, Ya.:
Bosonization of ZF algebras: Direction toward deformed Virasoro algebra.
J. Exp. Theor. Phys. {\bf 82}, pp. 1021-1045 (1996) }

\lref\BLZZ{Bazhanov, V.V., Lukyanov, S.L. and Zamolodchikov, A.B.:
Integrable structure of conformal field theory 
II. Q-operator and DDV equation. Commun. Math. Phys. {\bf190}, pp. 247-278
(1997)}

\lref\Colm{Coleman, S.: The quantum sine-Gordon
equation as the Massive Thirring model. Phys. Rev. {\bf D11},
pp. 2088-2097 (1975)}

\lref\SL{Lukyanov, S.: Low energy effective Hamiltonian
for  the XXZ spin chain. Nucl. Phys. {\bf B522}, pp.533-549   (1998) }

\lref\Mandelstam{Mandelstam, S.:
Soliton operators for the quantized sine-Gordon equation.
Phys.Rev. {\bf D11}, pp. 3026-3030 (1975)}

\lref\LZ{Lukyanov, S. and Zamolodchikov, A.:
Exact expectation values of local fields
in the quantum sine-Gordon model.
Nucl. Phys. {\bf B493}, pp. 571-587 (1997)}

\lref\tsve{
Controzzi, D., Essler, F.H.L. and Tsvelik, A.M.:
Dynamical properties of one-dimensional Mott insulators.
To appear in the proceedings of 
Newton Institute EuroConference 
on Strongly Correlated Electron Systems: Novel Physics and New
Materials, Cambridge, England, 26-30 June 2000  (cond-mat/0011439)}

\lref\gog{Gogolin, A.O., Nersesyan, A.A. and Tsvelik, A.M.:
Bosonization and Strongly Correlated Systems.
Cambridge University Press (1999)}

\lref\tsvee{ Essler, F.H.L. and Tsvelik, A.M.: Private communication}

\lref\LZS{Lukyanov, S. and Zamolodchikov, A.: To appear}

\lref\Fedya{Smirnov, F.A.: Form-factors in Completely
Integrable Models of
Quantum Field Theory. Singapore: World Scientific (1992)}

\lref\FZPar{
Fateev, V.A. and  Zamolodchikov, A.B.:
Parafermionic currents in the two-dimensional Conformal Quantum Field Theory
and self-dual critical points in $Z_N$ invariant statistical systems.
Zh. Eksp. Teor. Fiz. {\bf 89}, pp. 380-399 (1985)}

\lref\lik{Lukyanov, S.: Free field 
representation for massive integrable models.
Commun. Math. Phys. {\bf 167}, No.1, pp. 183-226  (1995) \semi
Form-factors of exponential fields
in the sine-Gordon model.
Mod. Phys. Lett. {\bf A12}, pp. 2543-2550 (1997)}

\lref\fatb{Baseilhac, P. and Fateev, V.A.:
Expectation values of local fields for a two-parameter family 
of integrable models
and related perturbed Conformal Field Theories.
Nucl. Phys. {\bf B532}, pp. 567-587 (1998)}

\lref\fatszz{Fateev, V., Lukyanov, S., Zamolodchikov, A. and
Zamolodchikov, Al.:
Expectation values of boundary fields in the boundary sine-Gordon model.
Phys. Lett. {\bf B406}, pp. 83-88 (1997)\semi
Expectation values of local fields in the 
Bullough-Dodd model and integrable perturbed
Conformal Field Theories.
Nucl. Phys. {\bf B516}, pp. 652-674 (1998)}
 
\lref\frad{Fateev, V., Fradkin, D., Lukyanov, S., Zamolodchikov, A. and
Zamolodchikov, Al.:
Expectation values of descendent fields in the sine-Gordon model.
Nucl. Phys. {\bf B540}, pp. 587-609 (1999) }

\lref\fatt{Fateev, V.A.:
Normalization factors in Conformal Field Theory and their applications.
Mod. Phys. Lett. {\bf A15}, pp. 259-270 (2000) }

\Title{\vbox{\baselineskip12pt
\hbox{RUNHETC 2001-05}}}
{\vbox{\centerline{Form factors of soliton-creating operators}
\centerline{ in the
sine-Gordon model}}}

\centerline{
Sergei Lukyanov and
Alexander Zamolodchikov}
\centerline{}
\centerline{Department of Physics and Astronomy,
Rutgers University}
\centerline{ Piscataway,
NJ 08855-0849, USA}
\centerline{and}
\centerline{L.D. Landau Institute for Theoretical Physics,}
\centerline{Chernogolovka, 142432, RUSSIA}

\centerline{}
\centerline{}
\centerline{}

\centerline{\bf Abstract}

\centerline{}

We propose explicit expressions for the form factors, including
their normalization constants, of topologically charged (or
soliton-creating)
operators in the sine-Gordon model. The normalization constants, which
constitute the main content of our proposal, allow one to find exact
relations between the short- and long-distance asymptotics of the
correlation functions. We make predictions concerning asymptotics
of fermion correlation functions in the massive Thirring model, 
$SU(2)$-Thirring model with anisotropy, 
and in the  half-filled Hubbard chain.

\centerline{}

\Date{February, 01}

\vfill
\eject

\newsec{Introduction}

Exact form factors are of significant interest in integrable
quantum field theories in two space-time dimensions, and much progress has 
been made during the last 25 years in computing these quantities
in various models, and in their applications to the analysis of 
correlation functions. By $n$-particle form factors we 
understand as usual the matrix elements
of any local (or ``semi-local'', see below) field operator ${\cal O}(x)$ 
between the vacuum and the states 
containing $n$ particles. One of the reasons for the interest lies
in the fact that
exact form factors allow one to generate large-distance expansions
for the correlation functions by inserting a complete set of states of 
asymptotic particles. If the factorizable
$S$-matrix of these particles is known, the form factors are determined 
by solving the ``form factor bootstrap equations'' (see\ \Fedya\ for coherent 
exposition of the bootstrap program. Refs.\oldref\ includes some early works on the
subject).

It is important
to note that these form factor bootstrap equations constitute a linear 
system, i.e. if a certain
collection of form factors solves the equations, the renormalized
collection (with all form factors associated with the same operator
${\cal O}$ multiplied by the same constant) also does; this of 
course represents the freedom in normalization of the field operator
${\cal O}$. On the other hand it is usually convenient to fix the 
normalizations of the field operators in terms of the short-distance
behavior of their correlation functions. If the short-distance behaviour 
is controlled by associated CFT, the two-point correlation
function of a spinless field ${\cal O}(x)$ 
has the asymptotic form
\eqn\qwe{\langle\,  {\cal O}(x) 
\,{\cal O}^{\dagger}(0)\, \rangle \to 
{1\over |x|^{4\Delta_{\cal O}}}\, .}
Conventional choice of the coefficient $1$ in the power law
\qwe\ fixes
(up to a phase, which usually can be fixed through other correlation 
functions) the normalization of ${\cal O}$. The problem arises of 
finding the specific normalization of the form factors  which corresponds
to the ``CFT normalization''\ \qwe.

In principle, this problem can be solved by analyzing the short-distance
behavior of the form factor expansion. However, in
practice it is usually possible
to compute only the first 
few terms of this expansion. While yielding in many
cases excellent numerical data even for rather small distances
(see e.g. \coloda), 
this truncated series does not provide exact analytic information
about the coefficient in the 
short-distance asymptotic\ \qwe. Therefore
the problem of determining the ``CFT normalizations'' of the 
form factors for a generic operator ${\cal O}$ remains largely 
open (notable exceptions being 
the cases when ${\cal O}$ is a component of
some conserved current, e.g. energy-momentum tensor, 
and the normalizations 
of its form factors can be fixed through the Ward identities, see e.g.
\Fedya).
Some progress was made in\ \LZ\  where the expectation values of 
{\it topologically neutral} primary 
operators in the sine-Gordon model (and in some other integrable QFT
\refs{\fatszz,\fatb,\fatt} ) were 
determined. This result makes it straightforward 
to fix the normalizations of
all higher form factors of these operators using  Smirnov's 
``annihilation pole'' relation\ \Fedya\ of the form factor bootstrap\ \lik.

In this paper we extend the above result to a class 
of {\it topologically charged} fields in the sine-Gordon model. 
Obviously, the vacuum expectation values of fields with nonzero topological 
charge vanish. The simplest form factor of an operator carrying topological
charge $n$ is its matrix element between the vacuum and an 
$n$-soliton (or $(-n)$-antisoliton, if $n$ is negative) state. 
In  Section 3 we propose an  explicit expression for
these simplest form factors of ``CFT-normalized'' primary 
operators. Our proposal, Eq.(3.1) below, actually concerns the
corresponding normalization factor ${\bf Z}_n$ in (2.12) (without
regard of normalization, some form factors with $n\not=0$ were
considered before in Refs.\drew). In Section 4 we carry out various checks
of this conjectured expression. 

Let us mention here that there is a
one-parameter family of primary fields 
of given topological charge $n$ (see
Sect.2 for details), and for generic values of the parameter $a$ these 
fields (denoted ${\cal O}_{a}^{n} $ henceforth) are not local (they are
``semi-local'' in the terminology of\ \FZPar).  
Nonetheless, there are several important reasons to be interested in 
form factors  and correlation functions of such nonlocal operators. Let us
mention a few. First, specific discrete values of the parameter $a$ provide
a set of local topologically charged fields 
which in fact coincide with the
fundamental Fermi fields and their 
composites in the massive Thirring model.
For other discrete values of this parameter these fields become components
of conserved nonlocal currents generating the affine quantum group symmetry
of the sine-Gordon model. These relations will be used in Sect.4. 
Second, analytic properties in the parameter $a$
are likely to be illuminating, as  was demonstrated in\ \fatszz\
for expectation 
values of neutral fields. We will say some words on this subject in
Sect.6. Finally, in many cases it has proved to be useful
to factorize local fields into nonlocal ones. For example, the 
$SU(2)$ Thirring model 
(and its anisotropic deformation) can be bosonized in 
terms of a sine-Gordon field plus a free massless boson, the correlation
functions of basic Fermi fields in this model being expressed through
the correlators of the nonlocal fields of the sine-Gordon theory. 
In Sect.5 we use this relation to make predictions about exact
asymptotics of fermion correlation functions in the $SU(2)$ Thirring
and Hubbard models.

\vfill
\eject

\newsec{ Topologically charged  fields in the sine-Gordon
model}

The sine-Gordon model, which is described by the 
euclidean action\foot{
Here and below we use the notations and conventions adopted in\ \LZ;
in particular, the language of
euclidean field theory is used by default.}
\eqn\ksiy{
{\cal A}_{sG} = \int d^2x\,
\bigg\{\,  {1\over{16\pi}}\,(\partial_{\nu}
\varphi)^2 -
2\mu\,\cos(\beta\varphi)\, \bigg\}\, , }
is invariant w.r.t. the field translations $\varphi(x) \to 
\varphi(x)\,\pm \,2\pi/\beta$. As a well known consequence, the model
admits a class of fields creating nonzero topological (or soliton)
charge. The simplest of these fields can be described as the ``dual
exponentials''
\eqn\yrtt{{\cal O}_{0}^{n}(x) = 
e^{{n\over{4\beta}}\,\int_{{\cal C}_x}\,\epsilon_{\mu\nu}\partial^{\mu}
\varphi\,  dx^{\nu}}\, ,}
with integer $n$. Here the integration goes along some contour ${\cal
C}_x$ (``Dirac string'') starting at infinity and ending at $x$, 
the precise shape of this
contour being very much arbitrary as the correlation functions
involving these fields depend only on its homotopy properties. The
field\  \yrtt\ creates a discontinuity of the field $\varphi$ along
${\cal C}_x$, so that the values of $\varphi$ right across this
contour differ by a constant equal $2\pi n/\beta$. In operator
formalism, the field operator associated with\ \yrtt\ creates the 
topological charge $n$. The fields\ \yrtt\ are mutually local and they 
have zero spin. Since the exponential in the r.h.s. of\ \yrtt\ requires
regularization this expression defines the field ${\cal O}_{0}^{n}$
only up to normalization. We fix this ambiguity by assuming the ``CFT
normalization'' \qwe, specifically
\eqn\slku{
\lim_{|x|\to 0}\,|x|^{{n^2}\over{8\beta^2}}\ 
\langle\,  {\cal O}_{0}^{n}(x) \,
{\cal O}_{0}^{n}(0)\,  \rangle = 1\,. }
Although the fields\ \yrtt\ are themselves local, they obviously are not 
local with respect to the field $\varphi(x)$. Nevertheless one can
define a class of useful nonlocal fields ${\cal O}_{a}^{n}(x)$ by 
``fusing'' the fields\ \yrtt\ with the exponentials of $\varphi$,
i.e. by taking the limits $\lim_{x' \to x}\ {\cal
O}_{0}^{n}(x)\exp(ia\varphi(x'))$. In view of the above nonlocality
this limit has a phase ambiguity, which can be fixed, for instance, by
specifying the direction, relative to the contour ${\cal C}_x$, from
which $x'$ approaches $x$. To eliminate all ambiguities one can choose
some cartesian coordinates $x=({\rm x, y})$, and define
\eqn\lkju{
{\cal O}_{a}^{n}(x) = \lim_{\epsilon\to +0} \,\exp\bigg\{-{n\over{4\beta}}\,
\int_{-\infty}^{{\rm x}}\,\partial_{\rm y}\varphi({\rm x, y})\,d{\rm
x}\bigg\}\,\exp\big\{\, ia\varphi({\rm x}+\epsilon, {\rm y})\, \big\}\ .}
In fact, rigid specification of the
contour ${\cal C}_x$ in\ \lkju\ results in inconveniently discontinuous
correlation functions involving these objects. To avoid this inessential
complication, in what follows by a correlation function
\eqn\lskdy{\langle\,  {\cal O}_{a_N}^{n_N}({\rm x}_N, {\rm y}_N)\cdots 
{\cal O}_{a_1}^{n_1}({\rm x}_1, {\rm y}_1)\, \rangle  }
we always understand the analytic continuation of this correlator from
the domain \break ${\rm y}_1 < {\rm y}_2 < 
\cdots < {\rm y}_N$. With this definition the 
correlation function\ \lskdy\ becomes a 
multivalued function of the coordinates
$x_1, \cdots, x_N$ which acquires a certain phase factor $\exp\big(-i\pi 
(a_j\,n_k + a_k\,n_j)/\beta\big)$ 
when the point $x_j$ is brought around $x_k$ counterclockwise.
In particular, the two-point correlation function can be written as
\eqn\uyt{\langle\,  {\cal O}_{a}^{n}(x)\,  {\cal
O}_{a'}^{-n}(0)\,\rangle = 
\Big(\, e^{i\pi}\, 
{{{\bar {\rm z}}}\over{ {\rm z}}}\, \Big)^{{{s(a',n)-s(a,n)\over 2}}}\ 
\ e^{i\pi\, {s(a,n)+s(a',n)\over 2}}\ \  {\cal
G}_{a,a'}^{(n)} (r)\,,}
where 
\eqn\kuytr{s(a,n) ={ a n\over \beta}\,,}
${\rm z}
= {\rm x}+i{\rm y}$, ${\bar {\rm z}} = {\rm x} - i{\rm y}$, and
the function 
${\cal G}_{a, a'}$ is real and depends only on the distance $r=\sqrt{
{\rm z}{\bar {\rm  z}}}$. 
In  the terminology of\ \FZPar\  the fields ${\cal O}_{a}^{n}$ are
``semi-local''. If $a\neq 0$ the field ${\cal O}_{a}^{n}$ carries
nonzero spin\ \kuytr. It becomes a local, bosonic or fermionic,
field when $s(a,n)$ takes integer or half-integer values. 
Note that the
definition\ \lkju\ fixes also the normalization of the field ${\cal
O}_{a}^{n}$, since the normalizations of ${\cal O}_{0}^{n}(x)$ and ${\cal
O}_{a}^{0}(x) \equiv e^{ia\varphi}$ are already fixed 
through\ \slku\ and\ \qwe, respectively. 
The same normalization can be specified
by the condition
\eqn\norm{{\cal G}_{a, - a}^{(n)}(r)\big|_{r \to 0} 
\to 
{1\over r^{2d(a,n)}}
\, \ \ \ \ \ \ \ \ \ \  {\rm with}\ \ 
\ \ \ \ d(a,n)=2 a^2+{n^2\over 8\beta^2}\ .}

The aim of this work is to describe the form factors of the 
operators ${\cal O}_{a}^{n}$ with $n$-soliton states. We will assume
the hamiltonian picture with coordinate ${\rm y}$ taken as euclidean
time. Note that in this picture the operators defined by\ \lkju\ obey
the simple hermiticity relation
\eqn\herm{\big({\cal O}_{a}^{n}\big)^{\dagger} = {\cal O}_{-a}^{-n} }
which allows us to restrict our attention to the form factors of ${\cal
O}_{a}^{n}$ with generic $a$ but $n \geq 0$.
We will use the notation
$A_{-}$ and $A_{+}$ for the soliton and
antisoliton; these particles carry negative
and positive units of the topological charge,
\eqn\lsdiai{H = {\beta\over{2\pi}}\,\int_{-\infty}^{\infty}\,\partial_{\rm
x}\varphi({\rm x, y}) \,d{\rm x}\, , }
respectively.
Conservation of the topological charge implies
that nonvanishing form factors of the operator ${\cal
O}_{a}^{n}$ are of the form
\eqn\higher{
\langle\, vac \mid {\cal O}_{a}^{n}(0)\mid A_{-}(\theta_{1})\cdots
A_{-}(\theta_{n+N})\,  
A_{+}(\theta_1')\cdots A_{+}(\theta_N')\, \rangle\ ,}
where $\theta_i$ and $\theta_j'$ denote rapidities of solitons and
antisolitons (for simplicity, here and below we ignore the possible
presence of breathers). Up to overall normalization, all these 
form factors can be written down in closed form, as certain 
$N$-fold integrals\ \lik. 
In the simplest case, $N=0$, an  explicit formula
exists,
\eqn\ffact{\langle\, vac \mid 
{\cal O}_{a}^{n}(0) \mid  A_{-}(\theta_1)\cdots
A_{-}(\theta_n)\, \rangle_{in}=
\sqrt{{\bf Z}_{n}(a)}\ e^{{i{\pi n a}\over
{2\beta}}}\ \prod_{m=1}^{n}\,e^{{a\over\beta}\theta_m}\,
\prod_{m<j}^{n}\,G(\theta_m - \theta_j)\ .}
Here
\eqn\lsdyy{G(\theta)=
i\ {\cal C}_1\ \sinh(\theta/2)\ \exp\Big(\int_{0}^{\infty}
{dt\over t}\ {\sinh^2 t(1-i\theta/\pi)\, \sinh(t (\xi-1))\over
\sinh(2t)\ \sinh(t\xi)\ \cosh(t)}\ \Big) }
with
\eqn\uio{{\cal C}_1=\exp\Big(-\int_{0}^{\infty}
{dt\over t}\ {\sinh^2 (t/2)\, \sinh(t (\xi-1))\over
\sinh(2t)\ \sinh(t\xi)\ \cosh(t)}\, \Big)\, , }
and we have assumed that the rapidities are arranged so that $\theta_1
< \theta_2 < \cdots <\theta_n$. Here and below we use
the notation
\eqn\uytt{\xi={\beta^2\over 1-\beta^2}\ .}
The only unknown component in\ \ffact\
is the real normalization constant ${\bf Z}_{n}(a)$. This constant controls
the long-distance asymptotics of the two-point function\ \uyt\ (with the
normalization already fixed by \norm) because it is
dominated by $n$-soliton intermediate states, namely
\eqn\liy{{\cal G}_{a,a'}^{(n)}(r) =
{\sqrt{{\bf Z}_n (a){\bf Z}_n (a')}\over n!}\,\int_{-\infty}^{+\infty}
\,\prod_{m<j}\,|G(\theta_m - \theta_j)|^2 \,\prod_{m=1}^{n}
e^{{(a-a')\over\beta}\theta_m
-Mr\,\cosh\theta_m}\,{{d\theta_m}\over{2\pi}} +\ldots\ ,}
where $M$ is the soliton mass, and the dots
stand for the subleading terms, which are of 
the order of $e^{-(n+2)Mr}$. Let us also stress here that once
the constant ${\bf Z}_n(a)$ in\ \ffact\ is fixed, 
the normalizations of all higher
form factors\ \higher\ are also fixed by the ``annihilation
pole'' condition of the form factor bootstrap (see\ \Fedya\ for details).
For example, using the approach of\ \lik, one can derive the following
expression for the $n+2$-particle form factor\ \higher,
\eqn\jaasdy{\eqalign{&
\langle\, vac\,|\, {\cal O}_a^{ n}(0)\, |\,
A_{-}(\theta_{1})\ldots A_{+}(\theta_{k})\ldots A_{-}(\theta_{n+2})\, 
\rangle_{in}=
{ i\, {\cal C}_2\sqrt{ {\bf Z}_n(a)}\over 4\, {\cal C}_1}
\ e^{{i\pi an\over 2\beta}}\
\prod_{m=1}^{n+2}
e^{ {a\over \beta} \theta_m}\times\cr&
e^{{\theta_k\over \xi}}\
\prod_{m<j} G(\theta_m-\theta_j)\ 
\bigg\{\, e^{{i\pi\over 2\beta^2}}
\, \int_{C_+}{d\gamma\over 2\pi} \, e^{-({2a\over \beta}
+{1\over \xi})\gamma}\
\prod_{p=1}^{k}W(\theta_p-\gamma)\prod_{p=k+1}^{n+2}
W(\gamma-\theta_p)-\cr &
e^{-{i\pi\over 2\beta^2}}
\, \int_{C_-}{d\gamma\over 2\pi} \,
e^{-({2a\over \beta}+{1\over \xi})\gamma}\
\prod_{p=1}^{k-1}W(\theta_p-\gamma)\prod_{p=k}^{n+2}
W(\gamma-\theta_p) \, \bigg\}\ .}}
Here the function $W$ and the constant\ ${\cal C}_2$\ are 
\eqn\lsdy{\eqalign{&W(\theta)=
-{2\over \cosh(\theta)}\ \exp\Big(-2\int_{0}^{\infty}
{dt\over t}\ {\sinh^2 t(1-i\theta/\pi)\, \sinh(t (\xi-1))\over
\sinh(2t)\ \sinh(t\xi)}\ \Big)\, ,\cr &
{\cal C}_2=\exp\Big(4\int_{0}^{\infty}
{dt\over t}\ {\sinh^2 (t/2)\, \sinh(t (\xi-1))\over
\sinh(2t)\ \sinh(t\xi)}\, \Big)\ .}}
The integration contours $C_{+}$ and $C_{-}$ in\ \jaasdy\ are
described as follows. 
The contour\ $C_{+}$\ starts 
from\ $-\infty$\ along the real axis of the complex\ $\gamma$\ plane, 
and  winds
around the poles of its integrand located in the strip\ $-{\pi\over
2}-0 < \Im m \gamma < {\pi\over 2} +0$, going first above the poles at 
$\theta_p+{i\pi\over 2},\ \
p=1,\ldots k$, and then below the poles at\ 
$\theta_p-{i\pi\over 2},\ \ p=k+1,\ldots n+2$,
and finally extends to\ $+\infty$\ along the real axis, as illustrated 
in Fig.1a. Similarly, the contour\ $C_{-}$\ goes above 
$\theta_p+{i\pi\over 2},\ \ p=1,\ldots k-1$ and below
$\theta_p-{i\pi\over 2},\ \ p=k,\ldots n+2$ of its integrand, see
Fig.1b. 
\midinsert

\centerline{\epsfbox{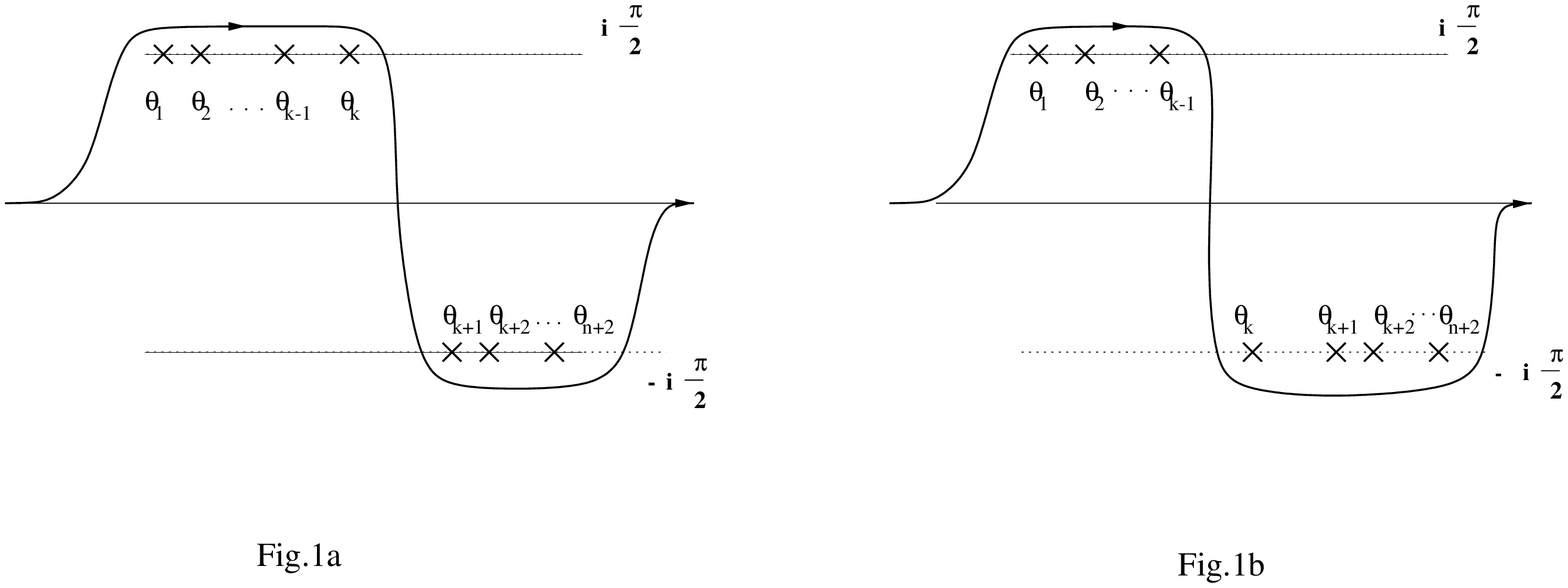}}
\vskip 7 truemm
\endinsert

\vfill
\eject

Under the assumed normalization of the fields\ ${\cal O}_{a}^{n}$, the
leading short-distance asymptotic of the two-point 
correlation function\ \uyt\ is
$$
{\cal G}_{a, a'}^{(n)}(r)\big|_{r \to 0}
\to { \langle\, e^{i(a+
a')\varphi}\, \rangle \over{r^{d(a,n)+d(a',n)-d(a+a',0)}}}\ ,$$
with the coefficient ${\cal G}_{a+a'} \equiv \langle\, e^{i(a+
a')\varphi}\, \rangle$ which can be read off from\ \LZ. To determine the
coefficient in its leading large-distance asymptotic\ \liy\ one
needs to know the normalization constant\ ${\bf Z}_{n}(a)$\ in\ \ffact.

\bigskip

\newsec{ Conjecture}

\vskip 0.1in

We propose the following explicit formula for 
the constant\ ${\bf Z}_n(a)$\ in\ \ffact:
\eqn\lsidiiu{\eqalign{&\sqrt{{\bf Z}_n(a)}=   \Big({{\cal C}_2
\over 2\,  {\cal C}_1^2}\Big)^{n\over 4}\ \ \Big(
{\xi\, {\cal C}_2\over 16}\, \Big)^{-{n^2\over 8}}\ \
\bigg[\, {\sqrt{\pi} M \Gamma({3\over 2}+
{\xi\over 2})\over \Gamma({\xi\over 2})}\, \bigg]^{d(a,n)}\times
\cr &
\exp\bigg[\, \int_{0}^{\infty} {dt\over t}\ 
\Big\{\, {\cosh(4\xi a t/\beta)\, e^{-(1+\xi) nt}-1\over 4\, \sinh(\xi t)
\sinh\big(\, (1+\xi) t\, \big)\, \cosh(t)}+{n\over 4\, \sinh(t\xi)}
-d(a,n)\ e^{-2 t}
\, \Big\}\, \bigg]\ .}}
Here the constants\ ${\cal C}_1$\ and\ ${\cal C}_2$\ are given by\ \uio\
and\ \lsdy, and\ $d(a,n)$\ is as in\ \norm. We would like to stress here
that this formula should be used only with\ $n \geq 0$.
For example, the form 
factor\ $\langle\,  vac\,  |\,  {\cal O}_{a}^{-n}(0)\,  |\, 
A_{+}(\theta_1) \cdots 
A_{+}(\theta_n)\,  \rangle_{in}$\ should
be written with\ $\sqrt{{\bf Z}_{n}(a)}$ 
(not $\sqrt{{\bf Z}_{-n}(a)}$) for its normalization constant.

In the next section some calculations supporting the
conjecture \ \lsidiiu\ are presented (and some are just mentioned).

\vskip 0.2in

\newsec{Supporting calculations}

\vskip 0.1in

\subsec{The case\ $n=0$}

In this case the field\ \lkju\ reduces to the exponential 
field\ $\exp(ia\varphi(x))$, and the form factor\ \ffact\ becomes its
vacuum-vacuum matrix element. Correspondingly, 
for\ $n=0$\ \lsidiiu\ reduces
to the expectation value of this exponential field (see\ \LZ).

\subsec{Free fermion point}

At\ $\beta^2 = 1/2$\ the sine-Gordon model reduces to free Dirac
fermions. In this case the form factors\ \ffact, and in particular
the normalization factors ${\bf Z}_{n}(a)$, can be calculated
directly\ \LZS. For this value of\ $\beta^2$ Eq.\lsidiiu\ coincides with
the result of\ \LZS.

\vskip 0.1in

\subsec{ Massive Thirring Model perturbation theory}

As is well known\ \refs{\Colm, \Mandelstam}\ the 
sine-Gordon theory is equivalent 
to the massive
Thirring model,
\eqn\mtm{{\cal A}_{MTM}=\int d^2x\
\Big\{\, {\bar \Psi}\gamma_{\mu}\partial_{\mu}
\Psi+{\cal M}\, {\bar \Psi}\Psi+{g\over 2}\,  J_{\mu}J_{\mu}\, \Big\}\ ,}
where $\Psi$ is a Dirac Fermi 
field,\ $J_{\mu}= {\bar\Psi}\gamma_{\mu}\Psi$\ is
its (non-anomalous) vector current, 
and the coupling constant\ $g$\ is related 
to\ $\beta$\ in\ \ksiy\ by
\eqn\gbeta{{g\over \pi} ={1\over 2\beta^2}-1\ .}
The action\ \mtm\ requires field and mass-term renormalizations, 
and the precise
relation between\ ${\cal M}$\ and\ $\mu$\ in\ \ksiy\ depends on the
renormalization scheme. The
fields $\Psi$, ${\bar\Psi}$ are related to ${\cal O}^{\pm 1 }_{\pm \beta/2}$ 
from  Sect.2 as follows\ \Mandelstam
\eqn\zpsi{\Psi (x) = {1\over\sqrt{2\pi\,  {\bf  Z}_\Psi}}\,
\bigg({{\cal O}^{\ 1}_{-\beta/2}\atop
{\cal O}^{\ 1}_{\beta/2}}\bigg)\, ,\ \ \ \  \quad
{\bar \Psi} (x)={1\over\sqrt{2\pi\, {\bf Z}_\Psi}}\ 
\Big(\, {\cal O}^{-1}_{-\beta/2}\, ,\, 
{\cal O}^{-1}_{\beta/2}\, \Big)\ ,}
where we have chosen a  chiral representation of the (euclidean) Dirac
matrices, i.e. 
$\gamma_{\rm x} = -\sigma_2$, $\gamma_{\rm y} = \sigma_1$. The 
constant ${\bf Z}_{\Psi}$ depends on the normalization
condition for the field $\Psi$. In what follows we assume the
most common renormalization scheme, in which the renormalized\ ${\cal M}$\ 
coincides with the soliton mass\ $M$, and the normalization of\ $\Psi$\ 
is fixed as follows,
$$
 (\, i\, {\not  p} +  M\, )\,S(p) \to 1 
\qquad {\rm as} \quad p^2 + M^2 \to 0\ .
$$
Here $S(p)$ is the momentum-space propagator,
$$
\langle\,  { \Psi}(x)\,  {\bar  \Psi}(0)\, \rangle =
\int\,S(p)\ e^{ipx}\ {{d^2p}\over{(2\pi)^2}}\ ,
$$
and\ ${\not  p}$\ stands for\ $\gamma_\mu p_\mu$. Comparing this equation 
with the\ $n=1$\ and\ $a=-a'=-\beta/2$ case of\ \liy,
one finds that this scheme corresponds to the choice\ 
$
{\bf Z}_{\Psi} = {\bf Z}_1 (\beta/2)
$\
in\ \zpsi. On the other hand, the short-distance behavior
of the correlation functions\ $\langle\,  {\cal O}^{1}_{-\beta/2}
{\cal O}^{-1}_{\beta/2}\, \rangle$\ is
given by\ \uyt,\norm. This leads to the following prediction for
the constant factor in the large-momentum asymptotic of the fermion 
propagator in this scheme,
\eqn\Sas{ S(p)\to -i\,  {\not p}
\ \ {M\pi \over
{\bf Z}_\Psi}\ {2^{1-2d_{\Psi}}\, \Gamma(3/2-d_{\Psi})\over 
\Gamma(1/2+d_{\Psi})}\
\ \ (p^2)^{d_{\Psi}-{3\over 2}}\,  \ \ \ \ \ \ \ {\rm as}\ \ \ 
p^2\to \infty\ .}
Here
$$
d_{\Psi} = {1\over{8\beta^2}}+{\beta^2\over 2}  
$$
and
\eqn\duiyu{\eqalign{& {\bf Z}_{\Psi}=\Big({{\cal C}_2
\over 2\,  {\cal C}_1^2}\Big)^{1\over 2}\ \ \Big(
{\xi\, {\cal C}_2\over 16}\, \Big)^{-{1\over 4}}\ 
\bigg[\, {\sqrt{\pi} M \Gamma({3\over 2}+
{\xi\over 2})\over \Gamma({\xi\over 2})}\, \bigg]^{2d_{\Psi}}\times\cr &
\exp\bigg[\, \int_{0}^{\infty} {dt\over t}\ 
\Big\{\, {\cosh(2\xi  t)\, e^{-(1+\xi) t}-1\over 2\, \sinh(\xi t)
\sinh\big(\, (1+\xi) t\, \big)\, \cosh(t)}+{1\over 2\, \sinh(t\xi)}
-2d_{\Psi}\ e^{-2 t}
\, \Big\}\, \bigg]\ .}}

This prediction can be verified in standard renormalized perturbation
theory in\ \mtm. Straightforward evaluation
of the diagram in Fig.2 yields
$$
S^{-1}(p) =i\,  {\not p} + M + \bigg({g\over{2\pi}}\bigg)^2\ \Sigma_2 (p)
+ O(g^3)\,,
$$
where 
\eqn\perturb{
\Sigma_2 (p) \to - i\,  {\not p}\ 
\bigg\{\, {\pi^2\over 3}-6
+\log\Big({p^2\over M^2}\Big)\, \bigg\}
\qquad {\rm as}\quad
p^2 \to \infty\ .}

\midinsert

\centerline{\epsfbox{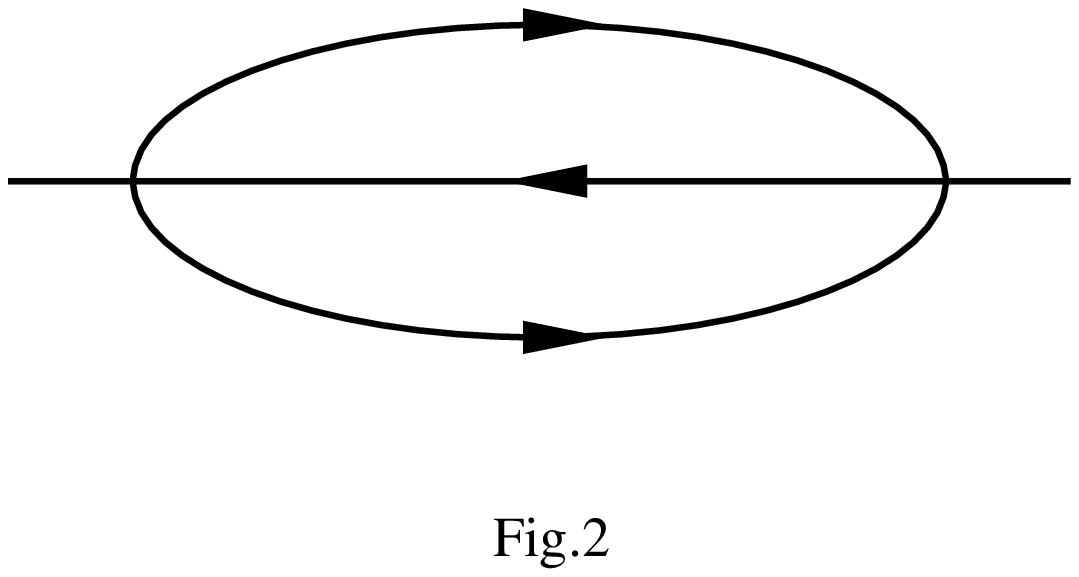}}
\endinsert

\noindent
Using a power expansion of\ \duiyu\ in the parameter\ \gbeta,
\eqn\lsid{{\bf Z}_\Psi={\pi M}\ \bigg\{\, 1+
\Big({g\over 2\pi}\Big)^2\ \Big(\, 
\log (M^2)+2\gamma_E+6-2\log 2-
{\pi^2\over 3} \, \Big)+O(g^3)\, \bigg\}\ ,}
where $\gamma_E=0.577216\ldots$ is Euler's constant, 
one can check that\ \Sas\ agrees with\ \perturb\ to
order\ $g^2$.

\subsec{Nonlocal Integrals of Motion}

As was mentioned in 
the Introduction, the fields\ ${\cal O}_{a}^{\pm 2}$\ with
appropriately chosen values of\ $a$\ coincide with the components of
the nonlocal currents found in\ \BeLeC. Namely, the fields
\eqn\gpm{\eqalign{&{\cal J}
_{\pm}(x) = {\cal O}_{\pm (2\beta)^{-1}}^{\pm 2}(x)\,, \quad
{\cal H}_{\pm}(x)=\pi\, \xi\, \mu  \ {\cal O}_{\pm\nu}^{\pm 2}(x)\ ,
\cr &
{\bar {\cal J}}_{\pm}(x) = {\cal O}_{\mp (2\beta)^{-1}}^{\pm 2}(x)\,, \quad
{\bar {\cal H}}_{\pm}(x)=\pi\, \xi\, \mu  \ {\cal O}_{\mp\nu}^{\pm 2}(x)
\ ,}}
with\ $\nu = (2\beta)^{-1}-\beta$, satisfy the continuity equations
\eqn\continuity{{\bar \partial}\,
{\cal J}_{\pm}(x) = \partial\,  {\cal H}_{\pm}\,, \qquad
\partial \,{\cal J}_{\pm}(x) = {\bar \partial}\,{\cal H}_{\pm}\ ,}
where $\partial ={1\over 2} ( \partial_{\rm x} - i\partial_{\rm y})\, ,
\ \ {\bar \partial}={1\over 2} (
\partial_{\rm x} + i\partial_{\rm y})$. They give rise
to four nonlocal integrals of motion
\eqn\xpm{\eqalign{&
Q_{\pm} ={1\over  {\bf Z}_Q}\   
\int_{-\infty}^{\infty}\,\big(\, {\cal J}_{\pm}({\rm x, y}) +{\cal H}_{\pm}
({\rm
x, y})\, \big)\,   d{\rm x}\ ,\cr &
{\bar Q}_{\pm} ={1\over  {\bf Z}_Q}\   \int_{-\infty}^{\infty}\,\big(\, 
{\bar {\cal J}}_{\pm}({\rm x, y}) + {\bar
{\cal H}}_{\pm}({\rm
x, y})\, \big)\, d{\rm x}\ ,}}
which generate 
(as was found in\ \BeLeC) the affine quantum group\ $U_q ({\widehat 
{sl(2)}})$
of level zero, with
$$
q=e^{i\pi/\beta^2}\,.
$$
More precisely, under a special choice of 
the ($\beta$- and $\mu$-dependent)
constant\ ${\bf Z}_Q$, 
the operators\ \xpm\ 
obey the commutation relations
\eqn\gtr{ Q_{-}\, {\bar Q}_+-q^2\  {\bar  Q}_+\,   Q_{-}=
{1-q^{2H}\over 1-q^{-2}}\, ,\ \ \ \ 
Q_{+}\, {\bar  Q}_--q^2\ {\bar Q}_-\,   Q_{+}=
{1-q^{-2H}\over 1- q^{-2}}\, ,}
where\ $H$\ is the topological charge\ \lsdiai.
The constant ${\bf Z}_Q$ can be expressed through the expectation
value of the fields\ $e^{i(\beta-1/\beta)\varphi}$,
\eqn\zbeta{{\bf Z}^2_Q = \mu \ \xi(1+\xi)\,\sin(\pi/\xi)\ \,
\langle\,  e^{i(\beta-1/\beta)\varphi}\, \rangle =
{1\over \Gamma^2(1+1/\xi)}
\bigg[\, {\sqrt{\pi} M \Gamma({3\over 2}+
{\xi\over 2})\over 2\, \Gamma(1+{\xi\over 2})}\, \bigg]^{{2\over \xi}}\, ,
}
where again\ $M$\ denotes the soliton mass.

The vacuum\ $|\, vac\,\rangle$\ is annihilated by all
the generators\ \xpm, while
the soliton-antisoliton pair\ $A_{\pm}$\ forms
two-dimensional representation of the algebra\ \gtr,
\eqn\actiono{\eqalign{&Q_{\pm}\, 
|\, A_{\pm}(\theta)\,
\rangle ={\bar Q}_{\pm}\, |\, A_{\pm}(\theta)\,\rangle=0
\, ,\cr
&Q_{\mp}\, |\, A_{\pm}(\theta)\,\rangle= e^{\theta\over \xi}\
|\, A_{\mp}(\theta)\,\rangle\, ,\ \ \ \  
{\bar Q}_{\mp}\, |\, A_{\pm}(\theta)\,\rangle=e^{-{\theta\over \xi}}\
|\, A_{\mp}(\theta)\,\rangle\,, }}
and $H\,|\, A_{\pm}(\theta)\, \rangle = 
\pm |\, A_{\pm}(\theta)\, \rangle$.
The action of these operators on any multisoliton
state is described by a 
standard\ $U_q({\widehat {sl(2)}})$\ coproduct, i.e.
\eqn\coprod{
\Delta(Q_{\pm}) = Q_{\pm} \otimes 1 + q^{\mp H}\otimes Q_{\pm}\, ,
\ \ \ \ \ \ \  
\Delta({\bar Q}_{\pm}) = {\bar Q}_{\pm}\otimes 1 + q^{\pm H}
\otimes {\bar Q}_{\pm}
\, .}
It is not difficult to see that consistency of the form factor 
formula\ \ffact\ with the above continuity equation\ \continuity\ and with
\actiono\ requires that the 
coefficient\ ${\bf Z}_n (a)$\ in\ \ffact\ satisfy
the following identities
$$
{\bf Z}_{2}\Big({1\over 2\beta}
\Big) =(\pi\xi\mu)^2 \  {\bf Z}_{2}\Big({1\over 2\beta}-\beta\Big) =\Big(
{M {\bf  Z}_Q\over 2\, {\cal  C}_1}\Big)^2\ ,
$$
which both are easily verified for\ \lsidiiu.

More interesting consistency conditions come from the following
observation. For all integer\ $n$\ and\ $m$\ the fields\ ${\cal
O}_{a_{n,m}}^{n}$ with
$$
a_{n,m} ={ n\over 4\beta} + {m\beta\over 2}
$$
are local with respect the currents\ ${\bar {\cal J}}_{\pm}$\ and
${\bar {\cal H}}_{\pm}$ defined in
\ \gpm\
(a similar set ${\cal O}^{n}_{-a_{n,m}}$ exists for the currents
${\cal J}_{\pm}$ and ${\cal H}_{\pm}$).
As a result, commutators (more precisely, ``$q$-commutators'') of
${\cal O}_{a_{n,m}}^{n}$ with the generators ${\bar Q}_{+}$ and 
${\bar Q}_{-}$ are expressed through the fields ${\cal O}_{a}^{n\pm
2}$ (with the parameter $a$  somewhat shifted) or their descendants.
In particular, one can derive the following commutation relation
\eqn\como{
{\bar Q}_+\ {\cal O}^{n}_{a_{n,1}}(x)-q^{n}\
{\cal O}^{n}_{a_{n,1}}(x)\ {\bar Q}_+=
{2\pi i\over {\bf Z}_Q}\  q^{{n\over
2}}\ 
{\cal O}^{n+2}_{a_{n-2,1}}(x)\ .}
This operator equation of course leads to certain relations between
the form factors. For instance, sandwiching\ \como\ between the vacuum 
and the state containing
$n+2$ antisolitons and using\
\coprod\ and\ \actiono\ to transform the matrix element appearing on
the left-hand side, one obtains
\eqn\gdtr{\eqalign{\sum_{k=1}^{n+2}\ q^{k-n-2}\
&e^{-{\theta_k\over\xi}}\ 
\langle\, vac\,  |\,  {\cal O}^{n}_{a_{n,1}}(0)\, |\, 
A_{-}(\theta_{1})\cdots A_{+}(\theta_k)\cdots A_{-}(\theta_{n+2})\,
\rangle_{in} =\cr &
-{2\pi i\over {\bf  Z}_Q}\  q^{-{n\over
2}}\
\langle\, vac\,  |
\,  {\cal O}^{n+2}_{a_{n-2,1}}(0)\, |\, A_{-}(\theta_{1})
\cdots A_{-}(\theta_{n+2})\, \rangle_{in}\ .}}
Now, substituting \ \jaasdy\ for the form factors in the left-hand side
of\ \gdtr, one observes that all but two integrals in the sum
cancel each other, which results in the relation
\eqn\ksdy{\eqalign{&{2\pi \over {\bf  Z}_q}\ 
{\sqrt{{\bf  Z}_{n+2}(a_{n-2,1})
\over {\bf  Z}_{n}(a_{n,1})}}=    
{i{\cal C}_2\over 4{\cal C}_1}\ \prod_{k=1}^{n+2}
 e^{ {\theta_k\over 2\beta^2}}\times\cr &
\bigg\{\, q^{{n\over 2}+1}\
\, \int_{C_+}{d\gamma\over 2\pi} \, e^{-{\gamma(n+2)\over 2\beta^2}}\ 
\prod_{k=1}^{n+2}W(\theta_k-\gamma)-
q^{-{n\over 2}-1}\
\, \int_{C_-}{d\gamma\over 2\pi} \, e^{-{\gamma(n+2)\over 2\beta^2}}\
\prod_{k=1}^{n+2}W(\gamma-\theta_k) \, \bigg\}\ ,}}
where $W$ is the function \ \lsdy. Thanks to the identity\
$W(\theta-i\pi)=W(-\theta-i\pi)$,
the sum of the integrals in the bracket\ $\big\{\, \cdots\, \big\}$
on the r.h.s. of\ \ksdy\ in fact reduces to the following limit
$$\Big\{\, \cdots\, \Big\}=\lim_{\Lambda\to +\infty}\,
\int_{-\Lambda-i\pi}^{-\Lambda+i\pi}\,  {d\gamma\over 2\pi}\ 
e^{-{\gamma(n+2)\over 2\beta^2}}\ 
\prod_{k=1}^{n+2}W(\theta_k-\gamma-i\pi)=i\ 
\Big({{\cal C}_2\xi\over 16}\Big)^{-{n\over 2}-1}\
\prod_{k=1}^{n+2}
e^{- {\theta_k\over 2\beta^2}}\  
  \ .$$
In turn, the limit can be evaluated using the asymptotic behavior
$$W(\theta-i\pi)\to \Big({{\cal C}_2\xi\over 16}\Big)^{-{1\over 2}}\ e^{-{
\theta\over 2\beta^2}}\ \ \ \ \ \ \ \ {\rm  as}\ \ \ \ \  \ 
\Re e\, \theta\to +\infty\,, $$
and therefore\ \como\ leads to the following relation between the
normalization factors in \ \ffact:
\eqn\aljjuy{ {{\bf Z}_{n+2}(a_{n-2,1})
\over {\bf Z}_{n}(a_{n,1})}=\Big({2 {\bf  Z}_Q 
 \over \pi\xi {\cal C}_1}\Big)^2\ 
\Big({{\cal C}_2\xi\over 16}\Big)^{-n}\ .}
It is possible to check that\ \lsidiiu\ indeed obeys\ \aljjuy.

\subsec{Scattering in the lattice XYZ model}

The sine-Gordon field theory describes the scaling limit of the integrable
$XYZ$ spin chain\ \Baxter,
\eqn\xyz{{\bf H}_{XYZ}=-{1\over 2}\ 
\sum_{k=1}^{N}\big(\, 
J_x\, \sigma_{k}^{x}
\sigma_{k+1}^{x}+J_y\, \sigma_{k}^{y}\sigma_{k+1}^{y}+J_z\, \,
\sigma_{k}^{z}\sigma_{k+1}^{z}\, \big)\ .}
As is known, at $J_y = J_x \geq |J_z|$ the spin chain\ \xyz\ is
critical, i.e. the
gap in the spectrum of\ \xyz\ vanishes, and its correlation length
$R_c$ becomes infinite in units of the lattice spacing\ \luth.
In the scaling limit $J_x - J_y \to 0$ the spin
correlations in\ \xyz\ are described by the field theory\ \ksiy\ with
\refs{\Jons,\luth}
$$
\eqalign{&\cos(\pi\beta^2) = J_{z}/J_{x}\ ,\cr & 
\bigg({{M\varepsilon}\over 4}\bigg)^{2-2\beta^2} \simeq 
{|J_{x}-J_{y}|\over 8 \sin^2(
\pi\beta^2)\, J_{x}} \to 0\ ,}
$$
provided
all relevant distances are $\sim R_c$, i.e. infinitely greater than
the lattice spacing $\varepsilon$. 
If, however, $J_{x}-J_{y}$ is small but finite,
there are corrections due to finite lattice size. These corrections 
can be taken into account by adding certain  irrelevant perturbations
to\ \ksiy, i.e. by using the ``effective action''
\eqn\effective{{\cal A}_{eff} = {\cal A}_{sg}+ 
\int \, {d^2 x\over 2\pi}\, \bigg\{\ 
{\lambda\over 2}\ \big(\, {\cal
O}_{0}^{4}(x) + {\cal O}_{0}^{-4}(x)\, \big)\, \varepsilon^{{2\over\xi}}-
\Big( \lambda_+\, {\bf T}{\bar {\bf T}}+
\lambda_- ( {\bf T}^2+
{\bar {\bf T}^2} )\, \Big)\, \varepsilon^2
+\ldots\,  \bigg\}\, ,}
where we have written down explicitly only the most significant of those
irrelevant terms. Here ${\bf T}$ and ${\bar {\bf T}}$ are components of the
sine-Gordon stress-energy tensor, in standard notations, and ${\cal
O}_{0}^{\pm 4}(x)$ are the fields\ \yrtt. In\ \effective\  the explicit
dependence on the lattice spacing $\varepsilon$ is exhibited to show
the relative smallness of various terms; the omitted terms are of
higher order in $\varepsilon$. Fortunately, the dimensionless coupling
constants $\lambda_{\pm}$ and $\lambda$ in\  \effective\ are known
exactly\ \SL, in particular
$$
\lambda ={4\ \Gamma(1+1/\xi)\over
\Gamma(-1/\xi)}\ \biggr(\,{ \Gamma\big(1+
{\xi\over 2}\big)
\over 2\, \sqrt\pi \ \Gamma\big({3\over 2}
+{\xi\over 2}\big)}\, \biggl)^{
{2\over \xi}}
\ .
$$
This makes\ \effective\ a working tool for determining the leading lattice
corrections to the scaling limit in the  $XYZ$ model.

Since the fields ${\cal O}_{0}^{\pm 4}$ carry nonzero topological
charge, the $\lambda$-terms in\ \effective\ generate processes violating
conservation of the topological charge, which is now conserved only
modulo 4. In particular, two solitons can be produced in
antisoliton-antisoliton scattering, with nonzero amplitude. In the Born
approximation this amplitude is
$$
S_{--}^{++}(\theta_2 - \theta_1) =
{(-i)\,\lambda\ \varepsilon^{{2\over\xi}}\over 4\pi\,M^2\, 
\sinh(\theta_2 -\theta_1)}\ \
_{out}\langle\, 
A_{+}(\theta_1) A_{+}\, (\theta_2)\,  |\, {\cal O}_{0}^{4}(0)\,  |\, 
A_{-}(\theta_1)\,  A_{-}(\theta_2)\,  \rangle_{in}\ ,
$$ 
where $\theta_1$ and $\theta_2$ ($\theta_1 < \theta_2$) are rapidities 
describing the  kinematics
of the scattering. The matrix element here can be obtained by crossing 
transformation  from the form factor\ \lsidiiu\ with 
$n=4$ and $a=0$. Simple calculation yields
\eqn\smatrix{S_{--}^{++}(\theta) = i\,  K \ \sinh(\theta/\xi)\ S(\theta)\ ,}
where 
$$S(\theta) = G(-\theta)/G(\theta) $$
is the sine-Gordon $S$-matrix element associated with elastic process
$A_{-}A_{-}\to A_{-}A_{-}$, and
\eqn\kone{K = -{\xi\, {\cal C}_1^2{\cal C}_2\,
\lambda\, \varepsilon^{{2\over\xi}}\over 16 \pi \, M^2}
\ \sqrt{{\bf Z}_4 (0)}\ .}
The appearance of this amplitude is of course expected. The scattering of 
excitations in the lattice model\ \xyz\ is described
by Baxter's elliptic S-matrix. It depends on the elliptic nome ${\tilde
p}$, whose precise
relation to the parameters in\ \xyz\ is rather transcendental (one can
consult\ \refs{\miw,\LYa}\ for 
details),  but here we only need to know that when
$|J_{x}-J_{y}| \to 0$ this nome also goes to zero as
$$
{\tilde p} \simeq \bigg({{M\varepsilon}\over 4}\bigg)^{2\over\xi}\ .
$$
If $J_{x}-J_{y} \neq 0$ this elliptic S-matrix has a  nonzero element
$S_{--}^{++}(\theta)$, presented explicitly in\ \refs{\miw,\LYa}.
One can check that 
when ${\tilde p}$ gets small this amplitude 
assumes precisely the form \smatrix\ with
\eqn\ktwo{K = 4\ {\tilde p}\,\sin(\pi/\xi)\, .}
This agrees with\ \kone\ provided
\eqn\zetfour{ \sqrt{{\bf Z}_4 (0)} ={M^2\over {\cal C}_1^2 {\cal C}_2}
\ \ {16\pi^2\, \xi\over
 \Gamma^2(1/\xi)}\
   \bigg({M \sqrt\pi\, \Gamma({3\over 2}
+{\xi\over 2})\over 2\, \Gamma(1+{\xi\over 2})}
\bigg)^{{2\over\xi}} \ .}
It is not difficult to verify that for given values of
the  parameters\ \lsidiiu\
indeed reduces to\ \zetfour.

\newsec{Deformed $SU(2)$-Thirring model}

The normalization factor\ \lsidiiu\ leads to certain predictions about
the asymptotic behavior of fermion correlation 
functions in the so-called 
deformed (or anisotropic) $SU(2)$ Thirring model. The latter is
described by the action (we use again euclidean notations)
\eqn\suthirring{{\cal A}_{DTM} =
\int\, d^2x\ \bigg\{\, \sum_{\sigma=\uparrow,\downarrow}\, 
{\bar\Psi}^{\sigma}
\gamma_{\mu}\partial_{\mu}\Psi_{\sigma} + {{g_0}\over
2}\ J_{\mu}J_{\mu} + {{g_{\parallel}}\over 2}\ J^{3}_{\mu}J^{3}_{\mu} 
+2\ {{g_{\perp}}}\,J^{+}_{\mu}J^{-}_{\mu}
\,  \bigg\}\ ,}
where $\Psi_{\uparrow}$, $\Psi_{\downarrow}$ is a doublet of Dirac
Fermi fields, and
\eqn\currs{J_{\mu} = {\bar\Psi}\gamma_{\mu}\Psi\,, \qquad J^{A}_{\mu} =
{\bar\Psi}\gamma_{\mu}\tau^{A}\Psi }
are their vector currents. The Pauli matrices $\tau^A = (\tau^3,
\tau^{+}, \tau^{-})$ in\ \currs\ act on the ``flavor'' indices
$\sigma=\, \uparrow, \downarrow$. The model\ \suthirring\ is
renormalizable, and
its coupling constants $g_{\parallel},\, g_{\perp}$ should be 
understood as ``running'' ones (the singlet coupling $g_0$ does not
renormalize). The corresponding RG flow pattern is known as 
Kosterlitz-Thouless flow\ (see e.g.\itdr).
In particular, in the domain $g_{\parallel} 
\geq |g_{\perp}|$ all RG
trajectories originate from the line $g_{\perp}=0$ of UV stable fixed
points, and\ \suthirring\ indeed defines a quantum field theory. 
In this domain (which is the only one that we discuss here) 
each RG trajectory is
uniquely characterized by the limiting value
$${\bar g}_{\parallel} = 
\lim_{L\to-\infty}\,g_{\parallel}(L)$$
of the running coupling
$g_{\parallel}(L)$ at extremely short distances ($L$ stands for the
logarithm of the length scale), i.e. the theory
\ \suthirring\ depends on two 
dimensionless parameters ${\bar g}_{\parallel}$ and $g_0$, besides the
mass scale appearing through dimensional transmutation. The 
model\ \suthirring\ attracts much attention in condensed matter 
physics, e.g. because of its relation to the scaling limit of the Hubbard
model\ (see e.g.\tsve).

As is well known\ (see e.g.\gog), 
the model\ \suthirring\ can be bosonized in terms of
the sine-Gordon field $\varphi(x)$, with $\beta$ in\ \ksiy\ related
to\ ${\bar g}_{\parallel}$\ by
\eqn\betag{{1\over \beta^2}=1+{{2\, {\bar g}_{\parallel}}\over\pi}\,,}
and a free massless boson $\omega(x)$. For the latter we assume
conventional CFT normalization 
$$\langle\,  \omega(x)\, 
\omega(0)\, \rangle = -2\, \log ({\rm z {\bar z}})\ .$$ 
The chiral components of the Fermi fields 
$$
\Psi_{\sigma}(x) =
{1\over\sqrt{2\pi}}\ \bigg({{\psi_{\sigma R}(x)}\atop
{\psi_{\sigma L}(x)}}\bigg)
$$ 
are expressed in terms of these bosons as
\eqn\fermions{\eqalign{&\psi_{ {\downarrow}L}  = 
\eta_{{\downarrow}}\, \ {\cal O}^{-1}_{-\beta/4}\ {\tilde{\cal
O}}^{\ 1}_{\gamma/4}\ ,\ \ \ \  \qquad 
\psi_{{{\downarrow}}R}  = \eta_{{\downarrow}}\ \,
{\cal O}^{-1}_{\beta/4}\ {\tilde{\cal O}}^{\ 1}_{-\gamma/4}\ , \cr & 
\psi_{{\uparrow}L}  = \eta_{\uparrow}\ \,
{\cal O}^{\ 1}_{\beta/4}\ {\tilde{\cal
O}}^{\ 1}_{\gamma/4}\ ,\ \ \ \ \ \,  \qquad \psi_{{\uparrow}R}  = 
\eta_{\uparrow}\, \
{\cal O}^{\ 1}_{-\beta/4}\ {\tilde{\cal O}}^{\ 1}_{-\gamma/4}\ ,}} 
where $\eta_{\sigma}=\eta_{\sigma}^{\dagger}$ are Klein factors
($\eta_{\uparrow}^2 = \eta_{\downarrow}^2 =1,\ 
\eta_{\uparrow}\, \eta_{\downarrow} = -
\eta_{\downarrow}\, \eta_{\uparrow}$).
In\ \fermions\ ${\cal O}^{\pm 1}_{a}$ are nonlocal fields\ \lkju\ of the
sine-Gordon part of the bosonic theory, and ${\tilde{\cal O}}^{\pm 1}_{a}$
are similar expressions in terms of the free field $\omega$, with $\beta$
replaced by $\gamma$, where
$$
{1\over \gamma^2} = 1+{{2\,  g_{0}}\over\pi}\ .
$$
These fields can also be written as 
\eqn\oomega{{\tilde{\cal O}}_{a}^{n}(x) =
\exp\Big\{\, i\, \Big(\, a-{n\over 4\gamma}\, \Big)\, \omega_R({\rm z})-
i\,   \Big(\,  a+{n\over 4\gamma}\, \Big)\, \omega_L({\bar{\rm z}})\, 
\Big\}\ ,}
through the right and left moving chiral parts of $\omega(x)$.
Note that each of the factors ${\cal O}$ and ${\tilde{\cal O}}$ in
\ \fermions\ is nonlocal (they each have spin $1\over 4$), while the
products appearing there are local Fermi fields. 

As a result of the factorization\ \fermions\ (known as spin-charge
separation in condensed matter applications) the correlation functions
of the fermions also factorize, for instance
\eqn\thcorr{
\langle\,  \psi
_{\sigma' R}(x)\,  \psi^{\dagger}_{\sigma
R}(0)\, \rangle =
{i\, \delta_{\sigma'\sigma}
\over z}\ \  \ r^{{1\over 2}- (\gamma-\gamma^{-1})^2/4}\ \ 
 {\cal G}^{(1)}_{-\beta/4, \beta/4}(r)\ 
   \ ,}
where again 
$
{\rm z} = {\rm x}+i\, {\rm y},\ r = \sqrt{\rm z {\bar z}}\, .
$ 
In writing\ \thcorr\ we have used\ \uyt, along with the well known form 
of the two-point 
correlation function of the free-field exponentials\ \oomega, 
$$
\langle\,  {\tilde{\cal O}}^{-n}_{a'}(x)\,  {\tilde{\cal O}}^{n}_{-a}(0)
\, \rangle =\Big(e^{i\pi}\, 
{{\bar {\rm z}}\over {\rm z}}\Big)^{{an\over\gamma}}\ 
\ r^{-4a^2-{n^2\over 4\gamma^2}}\ 
\ \delta_{a' a}\    .
$$
The short distance asymptotic form
of this correlation function follows from
\norm,
$$
\lim_{r\to 0}\  {{\rm z}\over r}\ r^{2d_\psi}\
\langle\,  \psi
_{\sigma' R}(x)\,  \psi^{\dagger}_{\sigma
R}(0)\, \rangle = i\  \delta_{\sigma'\sigma}\ ,
$$
with
$$
d_{\psi}={1\over 2}+{(\gamma-\gamma^{-1})^2\over 8}+
{(\beta-\beta^{-1})^2\over 8}\ .
$$
This of course is nothing but our convention concerning the
normalization of $\Psi_{\sigma}$ which we implicitly assumed in
writing\ \fermions. On the other hand, using\ \liy\ with $n=1$ one
can obtain its large-distance asymptotic,
\eqn\expass{\langle\,  \psi
_{\sigma' R}(x)\,  \psi^{\dagger}_{\sigma
R}(0)\, \rangle ={i\,  \delta_{\sigma'\sigma}\over
{\rm z}}\ \ 
{{{\bf Z}_{1}(\beta/4)}\over\sqrt{2\pi M}}\
\ \ r^{- (\gamma-\gamma^{-1})^2/4}\ e^{-M r} +
 O(e^{-3Mr})\ ,}
where\ $M$\ is the sine-Gordon soliton mass. While the functional form
of\ \expass\ is well known\ \tsvee,
the coefficient ${\bf Z}_1 (\beta/4)$
constitutes a  nontrivial prediction which follows from\ \lsidiiu.
Let us mention here that according to\ \expass, in the case $\beta =
\gamma = 1$ (which corresponds to the symmetric model\ \suthirring\ with
$g_0 =0$ and $g_{\parallel} = g_{\perp} > 0$) one has
\eqn\badass{ \langle\,  \psi
_{\sigma' R}(x)\,  \psi^{\dagger}_{\sigma
R}(0)\, \rangle 
\to {i\,  \delta_{\sigma'\sigma}\over {\rm z}}\ \times
\cases{   1\quad  \quad\ \ \ \ \ \ \ \      {\rm
as} \quad r \to 0\cr   C\ e^{-Mr}\ 
\ \ \ \ {\rm as} \quad r \to \infty }\ ,}
with
\eqn\CPAHOE{C={ \Gamma(1/4)\over 
2^{{5\over 4}}\ \sqrt{\pi}}\ \exp\bigg\{\, \int_0^{\infty}{dt\over t}\
{\sinh^2(t/2)\, e^{-t}\over \sinh(2t)\, \cosh(t)}\,
\bigg\}=0.921862\ldots\ .}
The numerical coefficients in the short- and long-distance
asymptotics in\ \badass\ are remarkably close; this suggests that in
this symmetric case the leading term in the long-distance asymptotic\
\badass\ may be rather good approximation for the correlation function
at all scales.

Eq.\ \badass\ can be translated to asymptotic formulae for
the fermion correlation functions in the half-filled Hubbard chain
\eqn\hubbard{{\bf H}_{Hub}=-t\sum_{j=-\infty}^{+\infty} \sum_{\sigma =
\uparrow, \downarrow} \big(\, c_{j ,\sigma}^\dagger
c_{j+1, \sigma} +
c_{j+1 ,\sigma}^\dagger
c_{j, \sigma}\, )+U \ \sum_{j=-\infty}^{+\infty}
\Big(\, n_{j,\uparrow}-{1\over 2}
\, \Big)\Big(\, n_{j,\downarrow}-{1\over 2}
\, \Big)\ .}
In\ \hubbard\ $c_{\sigma, j},\,  c_{\sigma, j}^{\dagger}$ are the 
Fermi operators,
$$\{\, c_{j ,\sigma}^\dagger\, ,c_{j' ,\sigma'}\, \}=
\delta_{\sigma\sigma'}\ \delta_{jj'}\ ,
$$
and $n_{j,\sigma}=c^{\dagger}_{j,\sigma}c_{j,\sigma}$.
As is known\ \refs{\fil,\lutem,\Mel,\Woynarovich,\Korep}, in
the symmetric case ($g_0 =0$ and $g_{\parallel} =
g_{\perp} > 0$), the field theory \ \suthirring\ describes the
scaling limit of this chain. If one sends $U \to +0$, the correlation
length
$$
R_c = {\pi\over 2}\ \sqrt{t\over U}\ e^{2\pi t\over U}
$$ 
diverges,
and the correlation functions of\ \hubbard\ at large lattice separations 
assume certain scaling forms. In particular, 
if\ $|j-j'|\gg 1$\ the equal-time fermion correlator
can be written as
\eqn\scdre{\langle\, c_{j' ,\sigma'}\, 
c_{j ,\sigma}^\dagger\, \rangle \to \delta_{\sigma'\sigma}\ 
{{\sin\big(\, {\pi\over
2}\, (j'-j)\, \big)}\over{\pi\,  (j'-j)}} \ \  F(\, |j'-j|/R_c\,)\ ,}
where $F(0)=1$. The scaling function $F(\tau)$ here is related
directly to the field-theoretic correlation function \ \thcorr\ with
$\beta=\gamma =1$. Therefore Eq.\badass\ leads to the following 
prediction for the
large-distance asymptotic of the scaling function $F$ in \ \scdre,
$$
F(\tau) =C\,e^{-\tau}+O(e^{-3 \tau})\, \ \
\ \ \ \ \ \ {\rm as}\ \ \ \tau\to\infty\ ,
$$ where $C$ is the constant\ \CPAHOE. 

\newsec{ Discussion}

Conjecture\ \lsidiiu\ for the factor ${\bf Z}_{n}(a)$ in\ \ffact\ is 
the main statement
of this paper. In certain sense it extends the 
proposal of\ \LZ\ about vacuum
expectation values of ${\cal O}_{a}^{0}(x) \equiv \exp(ia\varphi(x))$
to the case of fields ${\cal O}_{a}^{n}(x)$ carrying nonzero
topological charge $n$. In this connection we would like to add the
following remark.

As was shown in\ \fatszz, the vacuum expectation
values
$$
\langle\,  vac\,   |\,  e^{ia\varphi}\, |\, vac\, \rangle = 
\sqrt{ {\bf Z}_{0}(a)}
$$
obey the interesting analytic relation
\eqn\rr{\sqrt{ {\bf Z}_{0}(a)} = R_{0}(a)\ 
\sqrt{{\bf Z}_{0}(-{\cal Q}-a)}\,,}
where ${\cal Q}={1\over\beta}-\beta$, and the amplitude
\eqn\refo{R_{0}(a) = (\Lambda)^{-2{\cal Q}(2a+{\cal
Q})}\ \ {{\Gamma(1+{{2a}\over\beta}+{{\cal
Q}\over\beta})\, \Gamma(1-2a\beta-{\cal Q}\beta)}\over
{\Gamma(1-{{2a}\over\beta}-{{\cal
Q}\over\beta})\, \Gamma(1+2a\beta+{\cal Q}\beta)}}\  ,}
$$
\Lambda =
{\sqrt{\pi} M \Gamma({3\over 2}+
{\xi\over 2})\over2\,  \Gamma(1+{\xi\over 2})}\  
\Big({1+\xi\over \xi}\Big)^{\xi}
$$
can be related to the ``Liouville reflection amplitude''\ \ZZ\ by
analytic continuation $\beta \to ib$. Although our understanding of
the ``reflection relation''\ \rr\ is still far from being satisfactory, 
it has proved to be very useful in determining vacuum expectation values
of other fields in the sine-Gordon\ \frad\ and in other 
integrable models\ \refs{\fatszz,\fatb,\fatt}.
It is possible to
check that ${\bf Z}_{n}(a)$ in Eq.\lsidiiu\ satisfies a similar relation
\eqn\rnn{\sqrt{ {\bf Z}_{n}(a)} = R_{n}(a)\ \sqrt{
{\bf Z}_{n}(-{\cal Q} - a)}  }
with
\eqn\refn{
R_{n}(a) = (\Lambda)^{-2{\cal Q}(2a+{\cal
Q})}\ \ {{\Gamma(1+{{2a}\over\beta}+{{\cal
Q}\over\beta}+{n\over{2\beta^2}})\, \Gamma(1-2a\beta-{\cal Q}\beta +
n/2)}\over{\Gamma(1-{{2a}\over\beta}-{{\cal
Q}\over\beta}+{n\over{2\beta^2}})\, \Gamma(1+2a\beta+{\cal Q}\beta +
n/2)}}\ .}
At the moment we are not in possession of any clear interpretation of
the ``reflection amplitude''\ \refn, neither in terms of Liouville nor
any other CFT, nor even as a natural analytic interpolation of the
``normalization factors'' of\ \fatt\ (note that\ \refn\ does not have
$n\to -n$ symmetry, which the Coulomb-gas integrals defining the
corresponding normalization
factors clearly do, thus indicating a  significant ambiguity in such
analytic interpolation for $n\neq 0$). 
Nonetheless, Eq.\rnn\ could be helpful in the  search for generalizations 
of\ \lsidiiu.

\centerline{}

\centerline{\bf Acknowledgments}
 
\vskip0.5cm

The authors acknowledge helpful discussions with Alexei Tsvelik. 
They are also  indebted to  Chris Hooley for careful reading of 
the manuscript.
S.L.   heartily thanks the organizers
of the research program
``New Theoretical Approaches to
Strongly Correlated Systems'', at the Isaac Newton Institute
for Mathematical Sciences, Cambridge,
where parts of this work were done, for their
kind hospitality. 
He has especially   benefited  from discussions with 
Fabian Essler.

This  research
is supported in part by DOE grant \#DE-FG02-96ER10919.

\listrefs

\end